\documentclass[english,aps,preprint]{revtex4}
\usepackage[T1]{fontenc}
\usepackage[latin9]{inputenc}
\usepackage{float}
\usepackage{amsmath}
\usepackage{graphicx}
\usepackage{amssymb}

\makeatletter
\@ifundefined{textcolor}{}
{
 \definecolor{BLACK}{gray}{0}
 \definecolor{WHITE}{gray}{1}
 \definecolor{RED}{rgb}{1,0,0}
 \definecolor{GREEN}{rgb}{0,1,0}
 \definecolor{BLUE}{rgb}{0,0,1}
 \definecolor{CYAN}{cmyk}{1,0,0,0}
 \definecolor{MAGENTA}{cmyk}{0,1,0,0}
 \definecolor{YELLOW}{cmyk}{0,0,1,0}
 }

\makeatother

\usepackage{babel}

\begin{document}

\title{{\Large Applying quantum mechanics to macroscopic and mesoscopic
systems}}

\author{N. Poveda T.}
\email{nicanor.poveda@uptc.edu.co}
\affiliation{Universidad Pedagógica y Tecnológica de Colombia, Grupo de Física
Teórica y Computacional, Tunja (Boyacá), Colombia }

\author{N. Vera-Villamizar.}
\email{nelson.vera@uptc.edu.co}
\affiliation{Universidad Pedagógica y Tecnológica de Colombia, Grupo de Astrofísica
y Cosmología, Tunja (Boyacá), Colombia }

\begin{abstract}
There exists a paradigm in which Quantum Mechanics is an exclusively
developed theory to explain phenomena on a microscopic scale. As the
Planck's constant is extremely small, $h\sim10^{-34}$ $\mbox{J.s}$,
and as in the relation of de Broglie the wavelength is inversely proportional
to the momentum; for a mesoscopic or macroscopic object the Broglie
wavelength is very small, and consequently the undulatory behavior
of this object is undetectable. In this paper we show that with a
particle oscillating around its classical trajectory, the action is
an integer multiple of a quantum of action, $S=nh_{o}$. The quantum
of action, $h_{o}$, which plays a role equivalent to Planck's constant,
is a free parameter that must be determined and depends on the physical
system considered. For a mesoscopic and macroscopic system: $h_{o}\gg h$,
this allows us to describe these systems with the formalism of quantum
mechanics.
\end{abstract}

\maketitle

\section{Introduction}

With Fermat's principle, for the trajectory of an optical ray, the
eikonal equation of the geometric optics is obtained and with the
principle of stationary action (or principle of minimum action), for
the trajectory of particles, the Hamilton-Jacobi equation is obtained;
the two equations set up a link between geometrical optics and classical
mechanics. L. de Broglie postulated in 1924 that a particle can have
wave properties (duality wave-particle) \cite{De-Broglie}, the experimental
confirmation of this phenomenon led Schrödinger to postulate an equation
for the De Broglie associated wave, called Schrödinger equation, in
which a macroscopic scale could be reduced to the classical mechanics
of a particle \cite{Schrodinger}, this gave origin to undulatory
mechanics. With the collaboration of M. Born and P. Jordan, W. Heisenberg
\cite{Heisenberg} establishes an equivalent theory known as matrix
mechanics. Nowadays, both theories, undulatory and matrix mechanics,
are known by the generic name of quantum mechanics. From the beginning
quantum mechanics has been accompanied by much controversy, mainly
with respect to the conceptual meaning of the mathematical formulations,
leading to various interpretations (Copenhagen, statistics, and others),
nevertheless, there is no doubt of its validity for its effectiveness
in describe and predict various experimental results.\medskip

M. Born interpreted the square of the wavefunction as the probability
density of finding a particle \cite{Born}, and with this he was able
to overcome the rupture existing between matrix and undulatory mechanics;
in addition it allowed to establish an analogy between the temporary
evolution of the probability functions and those of a hydrodynamic
fluid. The hydrodynamic formulation of quantum mechanics was initiated
with E. Madelung \cite{Madelung}, which was later used by D. Bohm
\cite{Bohm}; in this formulation, the Schrödinger equation is reduced
to the Hamilton-Jacobi equation. The Hamilton-Jacobi equation shows
that particles of the ensemble are subject, not only, to a \textquotedbl{}potential
classic\textquotedbl{} but also to a \textquotedbl{}quantum potential\textquotedbl{}
that give rise to forces that determine the evolution of each particle.
The lack of knowledge about how to obtain the quantum potential has
blocked the development of the hydrodynamic formulation of the quantum
mechanics, for this reason it has been restricted to philosophical
and epistemological discussions, and to the interpretation of the
results obtained by numerical integration, however, were found some
applications in the area of physical chemistry \cite{Aplicaciones}.\medskip

There exists a paradigm in which Quantum Mechanics are an exclusively
developed theory to explain phenomena on a microscopic scale. As the
Planck's constant is extremely small, $h\sim10^{-34}$ $\mbox{J.s}$,
and as in the relation of de Broglie the wavelength is inversely proportional
to the momentum; for a mesoscopic or macroscopic object the Broglie
wavelength is very small, and consequently the undulatory behavior
of this object is undetectable. For this reason, in a mesoscopic or
macroscopic system, the quantum mechanics reduces to classical mechanics.
However, some have attempted to extend quantum phenomena at the mesoscopic
scale \cite{AMesoscopicos}\cite{QMacroscopica} and it has been shown
that some macroscopic systems like the Solar System \cite{ASolar}
and Extrasolar \cite{AExtra} have variables that are quantized, but
when trying to introduce quantum mechanics in these systems there
are difficulties due to the smallness of the Planck's constant.\medskip

In classical mechanics the trajectory of a particle is determined
by the principle of stationary action, $\delta S=0$. However, under
certain conditions, the particle may oscillate around its classical
trajectory quantizing the action, ie, the action is an integer multiple
of a quantum of action, $S=nh_{o}$. The quantum of action, $h_{o}$,
which plays a role equivalent to Planck's constant, is a free parameter
to be determined and depends on the physical system considered. For
a mesoscopic or macroscopic system, $h_{o}\gg h$, this allows us
to describe these systems with the formalism of quantum mechanics.
It is necessary to use, in large part, the Bohmian interpretation
\cite{Bohm} to give meaning to quantum mechanics on a mesoscopic
and macroscopic scale. However, to avoid the epistemological controversy,
we can simply consider that the mathematical formalism of quantum
mechanics is applicable to these scales, if we consider that the systems
have their own constant $h_{o}$, in this way, the quantum behavior
of a mesoscopic or macroscopic system can be given by the superposition
of different macroscopic states.

\section{Quantization of the action}

The Hamilton-Jacobi equation for a moving particle with mass $m_{o}$
(henceforth, the subscript \textquotedbl{}o\textquotedbl{} indicates
that this quantity is constant) under the action of a conservative
effective potential is given by:
\begin{equation}
H\left(q_{1},\nabla S;t\right)+\frac{\partial S}{\partial t}=0,\qquad H=\frac{\left(\nabla S\right)^{2}}{2m_{o}}+U\left(q_{1}\right),\label{eq:Hamilton-Jacobi}\end{equation}

\noindent where, $H$ is the hamiltonian; $S=S(r,t)$, the Hamilton's
principal function; $U=U(q_{1})$, the energy potential; $r=r(q_{1},q_{2},\ldots,q_{g})$,
is the trajectory of the particle in real space and $q_{i}=q_{i}(t)$,
$i=1,\ldots g,$ are the $g$ generalized coordinates, in function
of the parameter $t$ (time). If the total energy of the particle
is equal to the minimum non-zero value of the potential energy, $E_{o}=U_{o}=U(q_{o})$,
the particle follows a minimal trajectory, $r=r(q_{o},q_{2},\ldots,q_{g})$,
where $q_{o}=q_{1}(t)$. In consequence, the particle that follows
a minimal trajectory does not experience any force: $F=-\nabla U(q_{o})=0$;
it is equivalent to having a free particle. As in the hamiltonian
there are not explicitly generalized coordinates $q_{i}$ (cyclic
coordinates) and the time $t$, there are two conserved quantities:
linear momentum ($\nabla S\equiv p_{o}$) and total system energy
($-\partial S/\partial t=H\equiv E_{o}$). Therefore, the Hamilton's
principal function is separable: $S(r,t)=S_{r}(r)+S_{t}(t)=p_{o}\cdot r-E_{o}\cdot t$.\medskip

We transversally perturb the minimal trajectory of the particle making
the coordinate $q_{o}$ varies slightly, $q_{1}\simeq q_{o}$, i.e.,
$\left|\xi\right|=q_{1}-q_{o}$; we have $r$ and $q_{1}$ perpendicular.
If the potential is such that, $\left[\nabla^{2}U(q_{1})\right]_{q_{1}=q_{o}}=K_{o}$,
($K_{o}>0$), the particle experiences a restoring force in direction
opposite to the perturbation, so that$\left|\xi\right|\rightarrow0$.
Making an expansion in Taylor series of the potential energy, $U(q_{1})$,
around $q_{1}=q_{o}$, we can obtain:
\begin{equation}
E_{o}=\frac{p_{o}^{2}}{2m_{o}},\qquad E_{\xi}=\frac{p_{\xi}^{2}}{2m_{o}}+\frac{1}{2}K_{o}\xi^{2},\label{eq:Energias}
\end{equation}

\noindent el momentum $p_{o}=m_{o}\overset{\bullet}{r}=m_{o}v_{o}$
is always constant and tangential to the minimal trajectory $r=r(q_{o},q_{2},\ldots,q_{g})$,
while the momentum, $p_{\xi}=m_{o}\overset{\bullet}{\xi}$, is variable
and is along $q_{1}$; therefore, both momentum do not necessarily
point in the same direction, in this case they are orthogonal. The
total energy of the perturbed particle is: $E=E_{o}+E_{\xi}$, and
its total momentum: $p^{2}=p_{o}^{2}+p_{\xi}^{2}$. For clarity, in
(\ref{eq:Energias}) we have separated the terms for the energy of
the non-perturbed particle, $E_{o}$ (terms that depend on $r$) and
the corresponding term to the energy supplied by the perturbation,
$E_{\xi}$ (terms that depend on $q_{1}$). \medskip

With the Hamilton's canonical equation $\overset{\bullet}{p_{\xi}}=-\partial H/\partial\xi$
and $\omega_{o}^{2}=K_{o}/m_{o}$, we obtain the equation of a harmonic
oscillator $\overset{\bullet\bullet}{\xi(t)}+\omega_{o}^{2}\xi(t)=0$.
Making the change of variable $r=v_{o\xi}t$ and $k_{o}^{2}=\omega_{o}^{2}/v_{o\xi}^{2}$,
leads to the Helmholtz equation: $\nabla^{2}\xi(r)+k_{o}^{2}\xi(r)=0$.
The quantities $\omega_{o}=2\pi/T_{o}$ and $k_{o}=2\pi/\lambda_{o}$
correspond to the frequency and angular wave number, respectively.
Combining the two differential equations by $\psi=\psi(r,t)=\xi(r)\cdot\xi(t)$,
we get the wave equation: $\partial^{2}\psi/\partial t^{2}-v_{o\xi}^{2}\nabla^{2}\psi=0$,
whose solution is the function:
\begin{equation}
\psi=C\cdot\exp\left[i(k_{o}r-\omega_{o}t)\right],\label{eq:Funcion-onda}
\end{equation}

\noindent the coefficient $C$ s a complex constant. However, the function that
describes the oscillations must be real, $Re\left\{ \psi\right\} =\xi_{o}\cos(k_{o}.r-\omega_{o}\cdot t+\alpha_{o})$,
where $\xi_{o}=\sqrt{2E_{\xi}/K_{o}}$ is the amplitude of movement
and $\alpha_{o}$ is a phase constant that depends on initial conditions;
$v_{o\xi}=\omega_{o}\xi_{o}$, is the phase velocity of traveling
wave which does not necessarily coincide with the speed of the particle.
Therefore, the perturbation makes the particle oscillate harmonically
around the minimal trajectory, $r=r(q_{o},q_{2},\ldots,q_{g})$.\medskip

The minimal trajectory of the non-perturbed particle, $r=r(q_{o},q_{2},\ldots,q_{g})$,
covers a distance $L$ from a starting point, $r(q_{o},q_{2},\ldots,q_{g};t=0)$,
to an arrival point, $r(q_{o},q_{2},\ldots,q_{g};t=T)$. The particle
obeys Newton's second law, and accordingly its trajectory fulfills
with the principle of least action or principle of stationary action:
$\delta S_{r}(r)=\delta S_{t}(t)=0$. An additional condition for
the fulfillment of the principle of least action, is that the particle
must pass through the starting and arrival points, making it necessary
to impose the condition, $\xi(0)=\xi(T)=0$, the perturbed particle
travels a distance $L=n\lambda_{o}$ in a time $T=nT_{o}$, where
$n=1,2,\ldots$. Consequently, Hamilton's principal function is periodic
(in space and time): $S(r,t)=S(r+n\lambda_{o},t+nT_{o})$ and vanishes
for, $S(0,0)=S(L,T)=n(p_{o}\cdot\lambda_{o})-n(E_{o}\cdot T_{o})=0$;
terms within the parentheses correspond to a constant, which is denoted
by $h_{o}$. Note that the action is quantized, i.e., the action is
given by $n$ times a quantum of action $h_{o}$: $S_{r}(L)=nh_{o}$
or $S_{t}(T)=nh_{o}$, the quantum action does not correspond to Planck's
constant in Quantum Mechanics, it is only a free parameter that we
must determine and depends on the physical system in consideration.
When the perturbed particle travels a distance $L=\lambda_{o}$ is,
$h_{o}=S_{r}(\lambda_{o})$, or when a time passes $T=T_{o}$ is,
$h_{o}=S_{t}(T_{o})$; these expressions correspond to the de Broglie
and Planck relations, respectively:
\begin{equation}
p_{o}=\hbar_{o}k_{o},\qquad E_{o}=\hbar_{o}\omega_{o},\label{eq:Broglie-Planck}
\end{equation}

\noindent where, $\hbar_{o}=h_{o}/2\pi$, is the reduced quantum of action;
momentum and energy are given by a quantum of action. Note that if
we have a family of perturbed minimal trajectories that fulfill the
principle of stationary action: $\delta S_{r}(r)=\delta S_{t}(t)=0$,
the trajectories that additionally quantize the action are those where
the action is an integer multiple of the quantum of action $h_{o}$
and therefore, $\delta S_{r}(r)=\delta S_{t}(t)=\delta(nh_{o})=0$,
i.e., a trajectory that quantize the action also fulfills the principle
of stationary action.\medskip

If we have a physical system which is continuously perturbed and shows
trajectories that are repeated periodically, the particles which quantize
the action give origin to stationary waves (or stable trajectories)
that store the energy of the perturbation. These waves interfere constructively
giving rise to resonance phenomena or steady states (when the system
is perturbed with the resonance frequency). The particles that do
not quantize the action are represented by traveling waves which transport
the energy of the perturbance, these waves end up giving their energy
and allow us to explain the evolution of the system from a steady
state to another when the resonance frequency changes.

\section{\noindent Wave function and Schrödinger equation}

In Classical Mechanics, the wave function (\ref{eq:Funcion-onda})
does not correspond to anything observable, only the real part represents
a traveling plane wave that carries energy from the perturbation;
the term, $Re\{\psi(r,0)\}$, represents a sinusoidal plane-wave extended
along $r$ and the term, $Re\{\psi(0,t)\}$, causes it to move with
a phase velocity, $\omega_{o}/k_{o}=v_{o\xi}$, which does not match
the speed of the particle. Relations (\ref{eq:Broglie-Planck}) determine
the phase velocity $v_{o\xi}=\omega_{o}/k_{o}\equiv E_{o}/p_{o}$;
note that perturbed trajectories which quantize the action have a
determined phase velocity. In order to represent a particle using
equation (\ref{eq:Funcion-onda}), it is necessary to modulate the
amplitude of the wave with a positive real function, $A(r,t)=A(r-v_{o}t)$,
to limit the extent of the wave forming a wave packet and making the
covering moves without scattering with a group velocity, $v_{o}$:
\begin{equation}
\Psi(r,t)=C\cdot A(r,t)\cdot e^{\frac{i}{\hbar_{o}}S(r,t)}.\label{eq:Funcion-onda-S}
\end{equation}

The complex part of the equation (\ref{eq:Funcion-onda-S}) allows
taking into account the phenomena related to the waves superposition,
while the real part corresponds to the distribution function (or probability
density) of finding a particle at a given point: $\rho(r,t)=\left|\Psi\right|^{2}=\left|C\right|^{2}A^{2}(r,t)$.
To integrate above all the space the constant $C$ allows normalizing
the unit. The probability current is defined as $J=\rho v_{o}$. The
conservation of probability is given by the equation of continuity:
\begin{equation}
\frac{\partial\rho}{\partial t}+\nabla J=0,\label{eq:Continuidad}
\end{equation}

\noindent the probability density moves in space with the same speed
$\nabla S/m_{o}=p_{o}/m_{o}=v_{o}$ and the trajectory that travels
the particle. To the Hamilton-Jacobi equation (\ref{eq:Hamilton-Jacobi})
we add the term $U_{Q}=\frac{1}{2}K_{o}\xi^{2}$ which corresponds
to the quantum potential,
\begin{equation}
H+U_{Q}+\frac{\partial S}{\partial t}=0,\label{eq:Hamilton-Jacobi-Q}\end{equation}

\noindent this implies that the classical force is affected by another
force which generates the quantization of the action of physical system:
$F=-\nabla U-\nabla U_{Q}$. The quantum potential causes the particle
to oscillate around the classical trajectory which has an effect on
probability density, therefore it is necessary to establish a relationship
between $U_{Q}$ and $A$, which is given by the equation: $\nabla^{2}A+\frac{2mU_{Q}}{\hbar^{2}}A=0$;
by substituting this expression in the equation (\ref{eq:Hamilton-Jacobi-Q})
and with (\ref{eq:Continuidad}) we can construct the Schrödinger
equation \cite{Bohm}:
\begin{equation}
H\Psi\left(r,t\right)=i\hbar\frac{\partial}{\partial t}\Psi\left(r,t\right),\label{eq_Schrodinger}
\end{equation}

From the Hamilton-Jacobi equation (\ref{eq:Hamilton-Jacobi-Q}) we
obtain the trajectory followed by the particles and from Schrödinger
equation (\ref{eq_Schrodinger}) we obtain the distribution of particles
in the space or the probability density, $\rho(r,t)$.

\section{Applications}

\subsection{The Kundt tube }

Consider a horizontal cylinder, whose length $L_{o}$ is greater than
its radius, one end is closed and in the other one a source is placed
that emits sound waves at a certain frequency. Air particles with
a mass $m_{o}$ oscillate longitudinally within the tube, with respect
to their equilibrium positions, with an amplitude $\xi_{o}$. Standing
waves that form inside the tube have an angular frequency, $\omega_{n}=n\pi/T_{o}$,
and a wavenumber, $k_{n}=n\pi/L_{o}$, where $n=1,2,\ldots$. We can
obtain the quantum of action from the relation of de Broglie (\ref{eq:Broglie-Planck}):
$h_{o}=p_{n}\lambda_{n}=2\pi m_{o}v_{o}\xi_{o}$. The boundary conditions
establish the potential at which the particles obey: $U(r)=\left\{ \begin{array}{ccc}
0 & \mbox{sii} & 0<r<L_{o}\\
\infty & \mbox{sii} & 0\geq r\geq L_{o}\end{array}\right.$, substituting into the Schrödinger equation (\ref{eq_Schrodinger}),
we obtain: $\varphi_{n}(r)=\sqrt{2/L_{o}}\sin\left(k_{n}r\right)$
and $E_{n}=\hbar^{2}\pi^{2}n^{2}/2m_{o}L_{o}^{2}=m_{o}\omega_{n}^{2}\xi_{o}^{2}/2$.
Each of the physical systems studied in quantum mechanics have a classical
analog that represents it.

\subsection{The Solar System formation}

In the primitive solar nebula, the particles condense forming small
planetesimals of mass, $m_{o}$, which orbit the Sun. Through a process
of accretion, the planetesimals gave rise to the planets that form
the Solar System. As the mass of the sun, $M_{\odot}$, is extremely
large in comparison with the mass of the planetesimals, $M_{\odot}\gg m_{o}$,
the action of the Sun prevails and the interaction among them only
results in weak perturbations to their respective orbits.\medskip

The hamiltonian (in polar coordinates) for a planetesimal orbiting
the sun is, $\widehat{H}=-\hbar_{o}^{2}\nabla^{2}/2m_{o}-GM_{\odot}m_{o}/r$
($\nabla^{2}$ is the laplacian operator in spherical coordinates),
by substituting in the equation (\ref{eq_Schrodinger}) we obtain:
$E_{n}=-E_{o}/n^{2}$ $(n=1,2,\ldots)$, where $E_{o}=GM_{\odot}m_{o}/2a_{o}$
and $a_{o}=\hbar_{o}^{2}/GM_{\odot}m_{o}^{2}$. Of the radial solution,
\[
R_{n,l}\left(\rho\right)=\left[\frac{\left(n-l-1\right)!}{(n+l)!}\right]^{\frac{1}{2}}e^{-\rho/n}\left(\frac{2\rho}{n}\right)^{l}\frac{2}{n^{2}}\: L_{n-l-1}^{2l+1}\left(\frac{2\rho}{n}\right)\qquad(\rho=\frac{r}{a_{o}})\]

\noindent We obtain that the probability $\left|\rho R_{n,l}\left(\rho\right)\right|^{2}$
is maximum for $n=1,2,\ldots$ and $l=n-1$; and corresponds to the
Bohr radius: $r_{n}=a_{o}n^{2}$.\medskip

By replacing the current data for the average radius($r_{n}$) and
orbital speed ($v_{n}$) of each planet in the equation $v_{n}^{2}r_{n}=GM_{\odot}\simeq887,43$
$(\mbox{km/s})^{2}\mbox{au}$, the error percentual is obtained: Mercury
($0,1$\%), Venus ($0,1$\%), Earth ($0,1$\%), Mars ($0,4$\%), Ceres
($0,3$\%) (we have taken Ceres because it is the most massive of
the asteroid belt), Jupiter ($-0,1$\%), Saturn ($-0,6$\%), Uranus
($-1,0$\%), Neptune ($-1,6$\%) and Pluto ($1,8$\%). By taking the
average orbital distance of the Earth as the unit $r_{n} = 1$ $\mbox{au}$
and averaging the relation of $r_{n}$ with Venus $r_{n-1}$ and Mars
$r_{n+1}$ the quantum number is obtained for the Earth ($n=5$) and
approximate value of $a_{o}\simeq0,04$ $\mbox{au}$. The quantum numbers of
the remaining planets are obtained with $n=\sqrt{GM_{\odot}/a_{o}}/v_{n}$:
Mercury ($3$), Venus ($4$), Earth ($5$) and Jupiter ($11$). Imposing
the criterion that does not take into account planets with a percentage
error $>0,1$\% and taking the orbital radii ($r_{n}=a_{o}n^{2}$)
of these planets using a Chi-square fit we have: $a_{o}=0,04292\pm0,00038$
au ($\chi^{2}=0,00225$, $R^{2}=0.99956$, level of confidence
$95$\%). With this value we obtain the value of the reduced macroscopic
parameter $\hbar_{o}=m_{o}\cdot1,469\times10^{14}$ Js and the base
energy $E_{o}=m_{o}\cdot1,034\times10^{10}$ J. Note that the quantum
of action is re-scale in the accretion process so that the conformation
of the solar system is independent of the mass of the planetesimals.\medskip

The sequence of quantum numbers ($n$) for the Sun and each of the
planets is: $1$, $3$, $4$, $5$, $6$, $8$, $11$, $15$, $21$,
$26$, $30$, respectively. With the quantum number $n$ and the Bohr
radius $a_{o}$ we obtain the orbital radius, orbital speed and the
angular momentum of the planets: $L_{l}=\sqrt{l(l+1)}\hbar_{o}$ with
($l=n-1$). Not only magnitude of angular momentum is quantized but
also its orientation. The $\theta_{teo}$ angle between the angular
momentum and the $Z$ axis is calculated: $\theta_{teo}=\arccos(m_{l}/\sqrt{l(l+1})$.
The observed orbital inclination is given by the $\theta_{obs}$ angle
with respect to an axis $Z_{e}$ perpendicular to the plane of the
ecliptic, however, this axis is arbitrary, so there is a discrepancy
between the theoretical and the observed angle: $\theta_{teo}-\theta_{obs}=\theta_{o}$.
Minimizing this difference using a least squares fit is obtained,
$\theta_{o}=(26.05583\pm0.73604)^{o}$.\medskip

Based on data from orbital inclination we obtain $m_{l}$ for each
of the planets: $2$, $3$, $4$, $5$, $6$, $9$, $13$, $18$,
$23$, $21$. The Solar System in its beginning was subject to a process
of migration of jovian planets \cite{Niza}; considering that the
greater probability is given for $l=n-1$ y $m_{l}=l$, we have the
latest configuration of the Solar System as given by the sequence
of quantum numbers $n$: $1$, $3$, $4$, $5$, $6$, $7$, $10$,
$14$, $19$, $24$, $22$. In migration the quantum number $m_{l}$
is maintained while $n$ changes to the jovian planets.\medskip

\begin{figure}[H]
\begin{centering}
\includegraphics[scale=0.6]{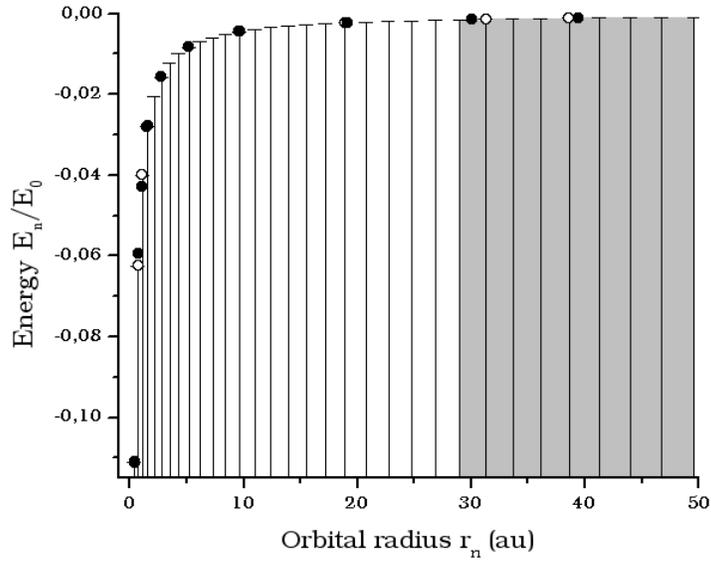}
\par\end{centering}

\caption{\label{fig:Energy}Relationship between energy levels and orbital
radii.}

\end{figure}

The Sun corresponds to $n=1$, consequently, a zero angular momentum.
As the filling of the orbitals due to the distribution of matter,
it appears that the existing matter in the quantum number $n=2$ has
been captured by the Sun, this could explain the inclination, the
differential rotation of the Sun and why its angular momentum corresponds
only to $\sim2\%$ of the whole Solar System. In Figure \ref{fig:Energy}
we observe that the continuum of energy levels ($\sim30-50$ $\mbox{au}$)
(shaded area) corresponds to the Kuiper Belt.

\section{Conclusions}

The primary goal of Classical Mechanics is to describe and explain
the motion of macroscopic objects affected by external forces. In
these systems the observable take continuous values obeying the Principle
of stationary action, nonetheless, the quantization of action that
just makes certain of this continuum values are permitted. It is clear
that Quantum Mechanics can not be applied on a macroscopic scale to
the entire physical system, only those where the action is quantized,
usually in resonance phenomena or systems where steady states appear.\medskip

There are many limitations in a microscopic scale when we want to
view or highlight certain theoretical or experimental facts, however,
it is possible to create a macroscopic mechanical model that is analogous
or equivalent to the microscopic quantum model, obtaining a \textquotedblleft{}toy-model\textquotedblright{}
where we can test some conjectures and observe the behavior and evolution
of the system under certain conditions. Although all of the above
mentioned has referred just to mechanic systems, it is possible to
extend these concepts to the study of another class of physical systems,
since in these systems the classical observables quantize or take
discrete values too.

\begin{acknowledgments}
The author would like to acknowledge the Dirección de Investigaciones
(DIN) of the Universidad Pedagógica y Tecnológica de Colombia (UPTC)
for its assistance.
\end{acknowledgments}

\end{document}